\documentclass[aps,pra,reprint,superscriptaddress,amsmath,amssymb]{revtex4-2}

\usepackage{graphicx}

\usepackage{listings}
\usepackage{longtable}
\usepackage{xcolor}
\usepackage{algorithm}
\usepackage[bookmarks,bookmarksnumbered,colorlinks,allcolors=blue]{hyperref}
\usepackage{bookmark}
\usepackage{ifthen}
\usepackage{multirow}
\usepackage{array}
\usepackage{siunitx}
\usepackage{qcircuit}
\usepackage[normalem]{ulem}

\newcolumntype{M}[1]{>{\centering\arraybackslash}m{#1}}

\definecolor{dkgreen}{rgb}{0,0.6,0}
\definecolor{gray}{rgb}{0.5,0.5,0.5}
\definecolor{mauve}{rgb}{0.58,0,0.82}

\newcommand{\figref}[1]{Fig.~\ref{#1}}

\begin{document}

\title{Millikelvin digital-to-analog converter for superconducting quantum processors}

\author{Ruizi Hu}
\thanks{These authors contributed equally to this work.}
\affiliation{S Lab, Quantum Science Center of Guangdong-Hong Kong-Macao Greater Bay Area, Shenzhen, China}

\author{Zongyuan Li}
\thanks{These authors contributed equally to this work.}

\author{Zhancheng Yao}
\thanks{These authors contributed equally to this work.}

\author{Yufei Wu}
\author{Qiang Zhang}
\author{Yining Jiao}
\author{Quan Guan}
\author{Lijing Jin}
\author{Wangwei Lan}
\author{Chengyao Li}
\author{Lu Ma}
\author{Liyong Mao}
\author{Huijuan Zhan}
\author{Ze Zhan}
\author{Ran Gao}
\author{Lijuan Hu}
\author{Kannan Lu}
\author{Xizheng Ma}
\author{Tenghui Wang}
\author{Peng Xiang}

\author{Chunqing Deng}
\email{dengchunqing@quantumsc.cn}

\author{Shasha Zhu}
\email{zhushasha@quantumsc.cn}
\affiliation{S Lab, Quantum Science Center of Guangdong-Hong Kong-Macao Greater Bay Area, Shenzhen, China}

\begin{abstract}
    Scaling superconducting quantum processors is increasingly constrained by the wiring, heat load, and calibration overhead associated with delivering high-resolution analog signals from room temperature to qubits at millikelvin temperature. Here we demonstrate a superconducting digital-to-analog converter (DAC) integrated with high-coherence fluxonium qubits in a multi-chip module architecture. The DACs generate persistent analog flux signals for tuning qubit parameters and are programmed deterministically using single-flux-quantum (SFQ) pulses, providing a digital interface compatible with established SFQ routing and demultiplexing technologies. Operating at millikelvin temperature, the DACs enable in-situ tuning of fluxonium qubits without measurable degradation of qubit coherence. The presented device provides a static control primitive for flux-tunable qubits, enabling parameter homogenization and eliminating the need for individual room-temperature DC bias lines. These results establish SFQ-programmable millikelvin DACs as a building block for digitally controlled superconducting quantum processors.
\end{abstract}

\maketitle

\section{Introduction}
Superconducting quantum processors have recently demonstrated rapid progress toward larger quantum systems, with several experimental platforms reaching the 100-qubit scale~\cite{kim2023evidence, Jin2025, Acharya2025, He2025}. However, implementing fault-tolerant quantum computing~\cite{Shor1995, Kitaev2003} for practical applications with superconducting technologies is expected to require processors containing more than one million physical qubits~\cite{Fowler2012, Reiher2017, gidney2025}. Scaling to such system sizes poses a fundamental challenge for the conventional control paradigm, in which each qubit is connected to room-temperature (RT) electronics through dedicated control lines. In this architecture, the number of control channels and associated RT electronics scales approximately linearly with the number of qubits, leading to an unsustainable growth in wiring complexity. As qubit counts increase, the sheer number of cables and their associated parasitic heat load will inevitably exceed both the cooling power and spatial capacity of standard dilution refrigerators~\cite{Krinner2019, Tian2025}. Consequently, integrating cryogenic control hardware directly at the millikelvin stage has emerged as a critical research frontier for enabling scalable superconducting quantum processors~\cite{Opremcak2018, Leonard2019, Acharya2023, Bao2024, Mukai2026}.

Among the various candidates for cryogenic controllers, superconducting digital logic based on single-flux-quantum (SFQ) pulses~\cite{McDermott2018} is particularly attractive due to its ultrafast switching speed, low power dissipation and natural compatibility with superconducting quantum circuits. Recent experiments have demonstrated that SFQ pulse sequences can directly drive coherent qubit gate operations~\cite{Liu2023}, and multiplexed SFQ architectures have further enabled routing of dynamic control signals to multiple qubits within a scalable control framework~\cite{Jordan2026}. These advances establish SFQ electronics as a promising platform for dynamic qubit control in superconducting quantum processors.

In addition to dynamic gate operations, superconducting qubits require precise static parameter tuning, such as flux biasing and frequency placement, to compensate for fabrication variations and to set optimal operating points. A promising approach for implementing static qubit control locally within the cryogenic environment is the use of superconducting digital-to-analog converters (DACs) that generate persistent analog flux signals directly on chip. DAC architectures of this type have been extensively developed for large-scale quantum annealing processors, where multiplexed addressing schemes program flux biases for thousands of qubits and couplers~\cite{Bunyk2014}. More recently, D-Wave has reported adapting this technology to fluxonium qubits for gate model quantum computing~\cite{DWave2025}, demonstrating the feasibility of integrating flux-biasing DACs with superconducting qubit circuits. However, in scalable superconducting quantum processors, the control system must therefore provide both high-speed dynamic gate operations and precise static parameter tuning, ideally implemented using compatible cryogenic electronics.

Here we demonstrate a millikelvin digital-to-analog converter integrated with superconducting quantum processors. The DAC is programmed deterministically using SFQ pulses, providing a digital interface compatible with established SFQ routing and demultiplexing technologies. We integrate the DAC with high-coherence fluxonium qubits in a multi-chip module architecture and operate the system below 20~mK. The DAC enables in-situ tuning of qubit parameters without measurable degradation of qubit coherence while eliminating the need for individual room-temperature DC bias lines. By combining static flux control with an SFQ-compatible programming interface, this work establishes millikelvin superconducting DACs as a control primitive for digitally addressable superconducting quantum processors.

\section{System Architecture and Device Overview}

\begin{figure}
    \includegraphics[width=1\linewidth]{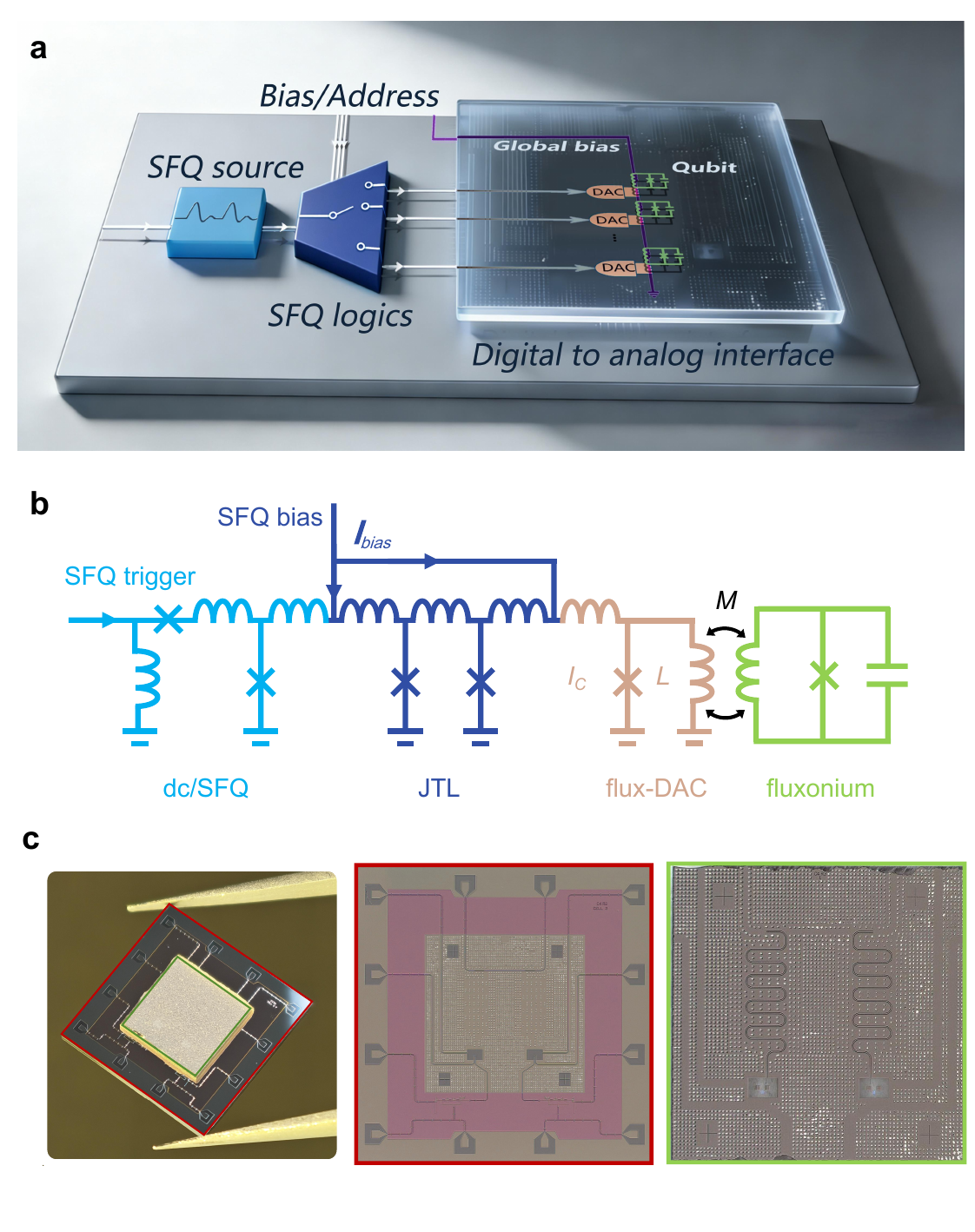}
    \caption{\textbf{SFQ-programmable DAC architecture and device implementation.} \textbf{a,} Conceptual architecture for scalable qubit control using SFQ-programmable digital-to-analog converters (DACs). \textbf{b,} Circuit-level implementation of the SFQ-programmable flux DAC integrated with a fluxonium qubit. SFQ pulses are generated by a dc/SFQ converter and transmitted through a Josephson transmission line (JTL) to the DAC. A bias current $I_\text{bias}$ sets the DAC operating point and provides bias for the dc/SFQ converter and JTL.
    \textbf{c,} Optical micrograph of the multi-chip module device (left), showing the integration of the qubit chip (right) and the digital control chip (middle).} \label{fig:arch}
\end{figure}

Figure~\ref{fig:arch}a illustrates the proposed architecture for scalable static control of superconducting qubits using SFQ-programmable DACs. SFQ pulses generated from a common source are processed by superconducting digital logic and routed through a demultiplexing network, enabling selective programming of individual DACs. Each DAC converts SFQ pulses into persistent flux stored in a superconducting loop, providing a programmable local flux bias for the qubit; such devices are hereafter referred to as flux DACs. This control architecture is fully compatible with mature SFQ digital logic circuits, allowing seamless integration with conventional SFQ signal conversion, transmission, and addressing schemes without reconstructing the core control framework~\cite{Likharev1991, Kirichenko2011}.

Each qubit is inductively coupled to a dedicated flux DAC that provides fine control of its operating point. A global bias line supplies a shared coarse flux offset across the qubit array and is also used to calibrate the DAC operating states. The final qubit operating point is therefore set by the combination of the global coarse bias and the dedicated DAC output for local fine tuning. In this architecture, control is naturally separated into a digital layer, responsible for SFQ pulse generation and routing, and a local analog layer, implemented by flux storage loops that generate persistent bias. Because the DAC states are stored persistently, qubit operating points can be configured once without continuous control signals or power dissipation, enabling scalable control without one-to-one wiring between room-temperature electronics and qubits.

To experimentally demonstrate this concept, we implement a prototype device consisting of a fluxonium qubit integrated with an SFQ-programmable flux DAC, as shown in Fig.~\ref{fig:arch}b. SFQ pulses are generated by a dc/SFQ converter and transmitted through a Josephson transmission line (JTL) to the DAC, where a Josephson junction with critical current $I_c$ and a storage inductor $L$ form the flux DAC. Each SFQ pulse injects a quantized flux into the storage loop, producing a persistent current that biases the qubit through a mutual inductance $M$ between the DAC and the qubit.

The device is realized in a multi-chip module (MCM) architecture, as shown in Fig.~\ref{fig:arch}c, in which the SFQ control circuitry and DAC are implemented on a separate chip from the qubit. This approach enables independent design and process optimization of high-coherence qubits and complex superconducting control circuits, allowing each to be fabricated and tested separately before integration.

\section{DAC Characterization}\label{app:Characterization}
\label{sec:DACchar}

\begin{figure}
    \includegraphics[width=1\linewidth]{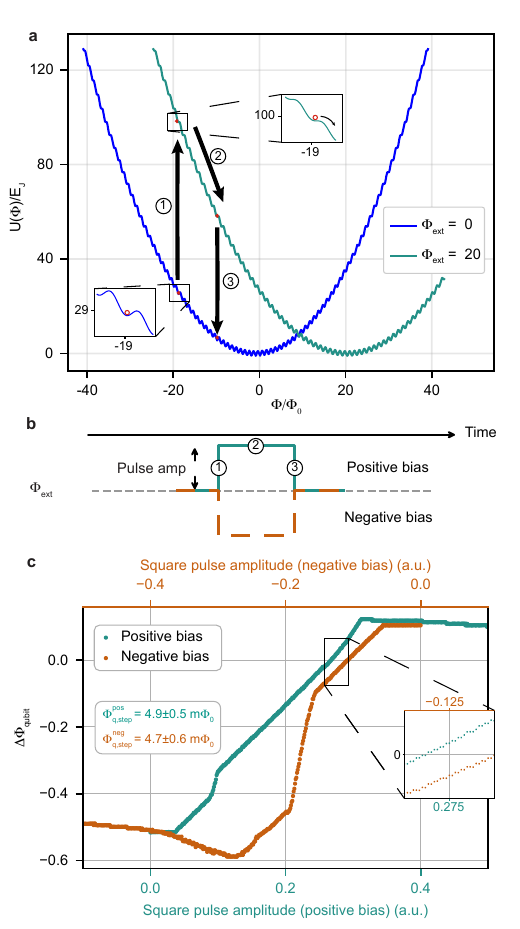}
    \caption{\textbf{DAC operation principal and characterization.} \textbf{a,} Potential energy against flux variable ($\Phi$) of a DAC with $\beta_L = 243$. Each local minimum corresponds to a metastable DAC state at $\Phi = N \Phi_0$, with $N$ an integer. External bias tilts the potential, enabling a controlled transition to the first available metastable state. The arrows indicate deterministic single-step transitions induced by programming pulses. \textbf{b,} Programming sequence of the DAC using square bias pulses. \textbf{c,} Measured flux increment $\Delta\Phi_{\mathrm{qubit}}$ as a function of pulse amplitude for positive and negative bias pulses, inferred from the qubit. Distinct plateaus indicate regimes where a single flux quantum is reliably added or removed, demonstrating deterministic, quantized control of the DAC state (insert). The asymmetry between positive and negative bias reflects device-specific characteristics of the flux trapping conditions.}
    \label{fig:DAC operation principal and characterization}
\end{figure}

The operation of the superconducting DAC is based on a flux storage mechanism implemented using an rf-SQUID, whose potential energy landscape supports multiple metastable states. The potential energy is given by~\cite{SQUIDHandbook2004}:
\begin{equation}
    U(\Phi) =E_J\left[\frac{1}{2\beta_L}\left(\frac{\Phi-\Phi_\mathrm{ext}}{\Phi_0}\right)^2-\cos\left(2\pi\frac{\Phi}{\Phi_0}\right)\right] \label{eq:potential energy}
\end{equation}
where $\Phi$ is the internal flux stored in the DAC loop and $\Phi_\mathrm{ext}$ is the externally applied flux bias controlled by $I_\mathrm{bias}$. For a large screening parameter ($\beta_L = 243$, corresponding to $I_c = 80~\mu\mathrm{A}$ and $L= 1~\mathrm{nH}$ in our device), the potential forms a tilted washboard landscape with multiple local minima, each corresponding to a metastable state at $\Phi = N \Phi_0$ (Fig.~\ref{fig:DAC operation principal and characterization}a). When the DAC resides in one of these minima, it stores a persistent current without requiring any bias, enabling non-volatile flux storage.

Programming the DAC corresponds to inducing controlled transitions between metastable states. This is achieved by applying a transient bias pulse that modifies $\Phi_\mathrm{ext}$, thereby tilting the potential landscape. When the tilt exceeds a critical threshold, the energy barrier to the nearest state is suppressed, allowing the system to transition to the first available metastable minimum. After the pulse is removed, the system relaxes and remains trapped in the new state, resulting in a discrete change of stored flux by approximately one flux quantum. In our experiment, this process is implemented using square bias pulses with tunable amplitude, as illustrated in Fig.~\ref{fig:DAC operation principal and characterization}b.

To quantitatively evaluate the operating range and precision of the DAC, we vary the amplitude of the programming pulses and monitor the resulting change in stored flux through qubit spectroscopy. Fig.~\ref{fig:DAC operation principal and characterization}c shows the measured flux increment $\Delta\Phi_\mathrm{qubit}$ as a function of pulse amplitude for both positive and negative bias directions. Clear plateaus are observed, indicating regimes where each pulse reliably induces a single-step transition between neighboring metastable states. These plateaus demonstrate deterministic, quantized control of the DAC state. From the plateau values, the minimum programmable step is measured to be $4.8 \pm 0.6 \,\mathrm{m}\Phi_0$.

The maximum output range of this DAC is approximately $0.7\,\Phi_0$, corresponding to about 146 discrete steps. In the ideal limit, this range is set by the stability condition of the rf-SQUID, which allows the stored current to vary within approximately $\pm I_c$. We further characterize multiple DACs, with their parameters summarized in Supplementary Material Sec.~S3. We note that the response curves for positive and negative bias exhibit asymmetry, as shown in Fig.~\ref{fig:DAC operation principal and characterization}c. This behavior arises from flux trapping and device-specific imperfections (see Supplemental Material Sec.~S1). Despite these imperfections, a well-defined overlap region remains in which the DAC response is monotonic and reproducibly controllable, providing a reliable operating regime as well as a reset mechanism for subsequent experiments.

\section{Compatibility with SFQ Control}\label{sec:SFQ control}
\begin{figure}
    \includegraphics[width=\linewidth]{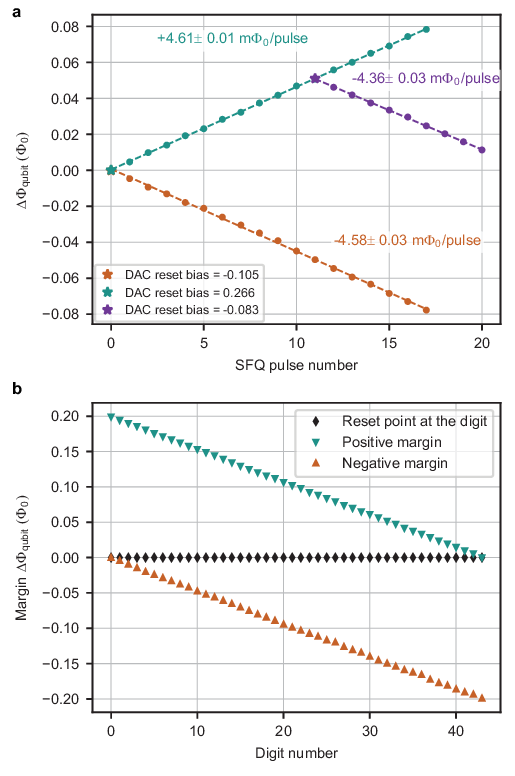} 
    \caption{\textbf{SFQ-programmed DAC operation.} \textbf{a,} DAC analog output as a function of the number of calibrated SFQ pulses. The DAC is initialized to different states using distinct bias values (stars; see legend). Subsequently, calibrated square pulses applied to the dc/SFQ converter generate SFQ pulses that increment or decrement the DAC state by one flux quantum (one digit) per pulse, resulting in a linear change of the DAC analog output (circles of corresponding colors). The direction of programming is controlled by the sign of the DAC bias and the polarity of the SFQ pulses, as indicated by the positive (green) and negative (orange and purple) slopes. \textbf{b,} Available SFQ programming margin of the DAC analog output as a function of DAC digit number within the linear operating regime (see Fig.~\ref{fig:DAC operation principal and characterization}c). The positive (green) and negative (orange) margins decrease approximately linearly with digit number, defining the accessible programming range around each initialized DAC digit (black markers) for a given bias.}
    \label{fig:SFQprogram}
\end{figure}

Building upon the DAC characterization, we next demonstrate its compatibility with SFQ-based digital control. We first initialize the DAC using an appropriate bias as characterized in Fig.~\ref{fig:DAC operation principal and characterization}c. The detailed initialization protocol is described in Supplemental Material Sec.~S5.

With the DAC initialized, calibrated square pulses are applied to the dc/SFQ converter, generating SFQ pulses. Each pulse deterministically updates the DAC state by one flux quantum, incrementing or decrementing the stored flux depending on the pulse polarity, resulting in a linear evolution of the DAC analog output (Fig.~\ref{fig:SFQprogram}a). In our device, each DAC step corresponds to a shift of approximately $4.58,\mathrm{m}\Phi_0$ in the qubit external flux.

The DAC can be programmed from any initial state using SFQ pulses, as illustrated in Fig.~\ref{fig:SFQprogram}a. While the pulse reset operations alone enable access to arbitrary DAC states, they require individual bias control for each DAC. To enable scalability, we instead use a global bias line shared across DACs, with programming performed via an SFQ demultiplexer tree (see Supplemental Material Sec.~S7). Combining arbitrary initial states set by the global bias with SFQ programming allows each DAC to reach a desired state, provided sufficient programming margin exists.

As shown in Fig.~\ref{fig:SFQprogram}b, the SFQ programming margin spans the linear operating regime. Starting from any reset point, additional positive or negative SFQ pulses can access higher or lower DAC states. This eliminates the need for individual bias lines for each DAC and provides a scalable control approach.

\section{Integration with a fluxonium qubit}\label{app:SFQ with qubit}

\begin{figure}
    \includegraphics[width = \linewidth]{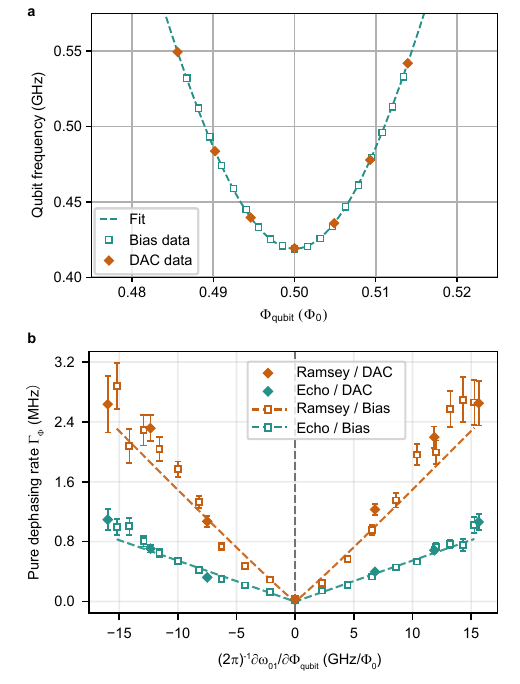}
    \caption{\textbf{Qubit characterization under DAC control.} \textbf{a,} Fluxonium qubit frequency as a function of external flux. Blue circles show spectroscopy measured using a conventional bias line, and the dashed curve is a fit to the fluxonium Hamiltonian, yielding $(E_C, E_J, E_L)/h = (1.3,5.08,0.806)~\text{GHz}$. Orange diamonds show the qubit frequencies obtained using the DAC analog output in combination with a fixed conventional bias. Comparing spectroscopy under DAC control with that obtained using conventional bias provides a precise calibration of the DAC analog output. \textbf{b,} Pure dephasing rate $\Gamma_\phi$ as a function of first-order flux sensitivity $\partial\omega_{01}/\partial\Phi_\mathrm{qubit}$. Green and orange squares denote Ramsey and echo data measured using the bias line, with corresponding fits to a $1/f$ flux noise model, yielding noise amplitudes $A_\Phi^\text{R} = 6.75 \pm 0.25\,\mu\Phi_0/\sqrt{\mathrm{Hz}}$ and $A_\Phi^\text{E} = 10.47 \pm 0.35\,\mu\Phi_0/\sqrt{\mathrm{Hz}}$. Green and orange diamonds show Ramsey and echo data measured under DAC control. Error bars represent the corresponding fit uncertainties.}
    \label{fig:qubit CT}
\end{figure}

The fluxonium qubit is first characterized using dedicated flux and charge control lines, while the DAC remains statically biased. As shown by the circles in Fig.~\ref{fig:qubit CT}, we measure the qubit transition frequency near its flux-insensitive point $\Phi_{\mathrm{ext}} = 0.5\Phi_0$, together with the dephasing times $T_2^{\mathrm{Ramsey}}$ and $T_2^{\mathrm{echo}}$ as a function of flux bias.

The qubit is then fixed at $\Phi_{\mathrm{ext}} = 0.5\Phi_0$ using the conventional flux line, after which the DAC is programmed via SFQ pulses as described in Sec.~\ref{sec:SFQ control}. The DAC programming sequence is performed once prior to repeated qubit operations, including initialization, gate execution, and readout.

The resulting flux shifts induced by the DAC are indicated by the diamond markers in Fig.~\ref{fig:qubit CT}a. Each DAC state change produces a flux shift of $4.58 \pm 0.01\,\mathrm{m}\Phi_0$, consistent with the calibrated DAC response.

To evaluate the impact of DAC operation on qubit coherence, we compare dephasing measurements obtained using the dedicated flux line with those obtained under combined flux line and DAC control. The data are analyzed using a $1/f$ flux noise model~\cite{Yoshihara2006, Bylander2011}, yielding flux noise amplitudes $A_\Phi^\text{R} = 6.75 \pm 0.25\,\mu\Phi_0$ and $A_\Phi^\text{E} = 10.47 \pm 0.35\,\mu\Phi_0$ from Ramsey and echo measurements, respectively. These values are consistent with reported fluxonium noise levels in the literature without on-chip DAC integration~\cite{Sun2023, Fei2025, Azar2026}. We further observe a largely flux-independent energy relaxation time with an average $T_1 = 82\pm17\,\mu\mathrm{s}$ across the measured bias range, independent of the DAC state (see Supplemental Material Sec.~S6). 
The dephasing and relaxation rates measured with DAC control agree well with those obtained using conventional bias. This agreement indicates that varying the DAC persistent current—thereby biasing the qubit through DAC state changes—does not introduce additional dephasing or relaxation, demonstrating compatibility with high-coherence fluxonium operation.

\section{Conclusion and outlook}

The demonstrated DAC provides a static control primitive that is naturally compatible with superconducting digital control architectures based on SFQ logic. In contrast to conventional flux-bias lines driven by room-temperature electronics, the DAC is programmed directly by SFQ pulses, enabling seamless integration with cryogenic digital routing networks.

The scalable programming architecture, initially presented in \figref{fig:arch}a, is depicted with greater detail in Supplemental Material Fig.~S6. In this scheme, an SFQ pulse generator feeds a binary DEMUX tree composed of cascaded 1:2 routing elements. Each stage of the tree is controlled by a select line, such that an input pulse is deterministically routed to one of $n$ outputs after passing through $\log_2(n)$ layers. As a result, a system with $n$ DACs can be addressed using only $\log_2(n)$ control lines, each operating at low bandwidth compared to the SFQ pulse stream. At the output of the tree, each branch connects to a DAC, while a shared global bias line provides the energy required for flux insertion. Only the selected DAC receives the complete signal path and is incremented by one flux quantum, while all other DACs remain unaffected. This architecture can be further generalized using shift-register-based addressing schemes~\cite{Kaplan1995}, in which SFQ pulses are sequentially loaded into a register chain that selects a target DAC. In this approach, the number of external control lines can be reduced to a constant independent of system size, at the expense of increased programming latency. 

Beyond static frequency tuning, the demonstrated DAC can be extended to control the amplitude of microwave drive signals through integration with tunable coupling elements. For example, a DAC-controlled tunable coupler can modulate the effective coupling between a qubit and a drive line or resonator, enabling programmable control of drive strength at the qubit. In this way, the DAC provides a general mechanism for in-situ calibration of both qubit frequencies and control amplitudes using cryogenic digital electronics.

This capability has important implications for large-scale superconducting quantum processors. Fabrication-induced variations lead to significant qubit nonuniformity, which remains a key limitation for scalable architectures. By enabling local, programmable adjustment of qubit parameters and control amplitudes, the DAC provides a means to homogenize qubit behavior across large arrays. Such homogenization is essential for implementing parallel and multiplexed gate operations~\cite{Zhao2024}, in which many qubits are driven simultaneously using shared microwave control signals. This addresses a central disadvantage of superconducting qubits, namely the variability introduced during fabrication.

From a system perspective, the combination of SFQ-based dynamic control and DAC-based static configuration suggests a pathway toward fully digital cryogenic control stacks, in which both fast gate operations and slow calibration functions are implemented within a unified superconducting electronics platform. The static nature of the DAC further ensures that this control paradigm incurs negligible steady-state power dissipation, as energy is consumed only during infrequent programming events. This characteristic is particularly important for operation at millikelvin temperatures and supports the scalability of the approach to large qubit systems.

Future work may explore the integration of large DAC arrays with SFQ routing networks, multiplexed addressing schemes, and co-design with qubit architectures, with the goal of realizing densely integrated control systems that overcome wiring and variability challenges in superconducting quantum processors.

\begin{acknowledgments}
This research is supported by the Guangdong Provincial Quantum Science Strategic Initiative (Grant No. GDZX2503003). The authors would like to express their gratitude to the Westlake Center for Micro/Nano Fabrication, Synergetic Extreme Condition User Facility (SECUF), Zhejiang QizhenTek Co. Ltd., and Micro-Nano Frbrication and Device Center in the Songshan Lake Materials Laboratory for their essential technical support during the chip fabrication process. We thank Dr. Chao Zhang and Dr. Zhen Yang from Instrumentation and Service Center for Physical Sciences at Westlake University for the supporting in device characterization. The ideas of this project were initiated in Z-Axis Quantum in early 2024. We thank Dr. Yaoyun Shi for his participation in the planning of the project, and his contribution to securing the necessary resources.
\end{acknowledgments}

\bibliography{refs}

\end{document}


\title{Supplemental Material for\\``Millikelvin digital-to-analog converter for superconducting quantum processors''}

\maketitle

\section{DAC design}\label{app:design}
\begin{figure}
    \includegraphics[width = \linewidth]{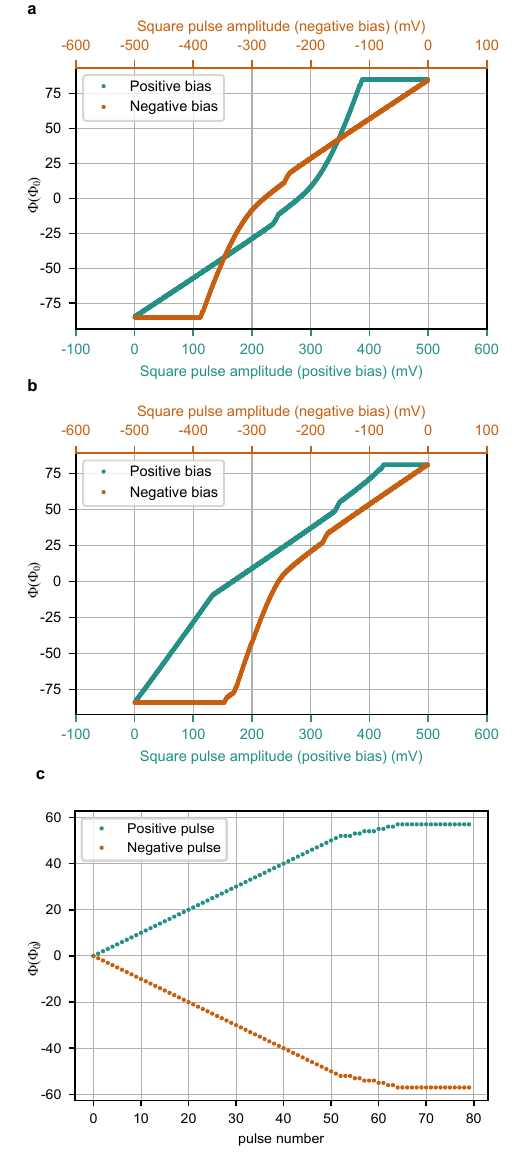} 
    \caption{\textbf{Numerical simulation of DAC characteristics.} 
    \textbf{a}, Stored flux $\Phi$ as a function of square pulse amplitude under positive and negative bias conditions, following initialization with a large reverse pulse that saturates the DAC to its minimum or maximum state.
    \textbf{b,} $\Phi$ as a function of square pulse amplitude in the presence of flux trapping in the superconducting loop containing the DAC and the bias current path. The resulting asymmetric response between positive and negative bias reproduces the behavior observed in Fig.~2c of the main text.
    \textbf{c,} $\Phi$ as a function of the number of applied SFQ pulses under a DAC bias current of $0.9 I_c$, for both positive and negative pulse polarities.}
    \label{fig:dac_design}
\end{figure}

The structure of the proposed DAC circuit is illustrated in Fig.~1b in the main text, and simulations are conducted via Josim, with the corresponding results presented in \figref{fig:dac_design}.To identify the inherent operating range of the standalone DAC, independent simulations are implemented by directly applying a square wave to the bias terminal, and the obtained results are shown in \figref{fig:dac_design}a. Results indicate that the adjustable range of the standalone DAC is considerably wider than that under the dc/SFQ-driven integrated operating mode. The slope variation in the curve of \figref{fig:dac_design}a stems from the prior switch of the front-end SFQ circuits relative to the Josephson junctions (JJs) inside the DAC under specific bias conditions. This timing difference in device response directly leads to the slope change in the curve, and further clarifies the fundamental cause of the restricted adjustable range in the integrated working mode.

In addition, we simulated the DAC performance with flux trapping in the loop consisting of the JJ in the dc/SFQ, the inductor in the SFQ bias, and the JJ in the flux-DAC. The result is shown in \figref{fig:dac_design}b. The flux trap significantly degrades the symmetry of the DAC under positive and negative pulse modulation, which is consistent with our test results.

To characterize the practical operating performance of the DAC integrated with SFQ logic circuits, simulation tests are performed under dc/SFQ driving, and the relevant results are displayed in \figref{fig:dac_design}c. The nonlinear region observed arises from the magnetic flux stored in the DAC, which interferes with the front-end SFQ circuit and impedes the normal transmission of SFQ pulses. This result demonstrates that the DAC reduces the operating margin of the front-end SFQ circuit, thereby narrowing its own adjustable range.

\section{Fabrication process}\label{app:fab}

\subsection{Digital control chip}
The digital control chip utilizes a three-metal layers superconducting digital architecture on a high-resistivity (\SI{>10}{\kilo\ohm\cdot\centi\metre}) \SI{4}{inch} Si (100) substrate. While the fabrication follows the general framework described in Ref.~\cite{SIMIT2021, AIST2023}, the following strategic modifications were implemented to optimize performance:
\begin{itemize}
    \item \textbf{Critical Current Control:} Overlapping Al/AlO$_x$/Al Josephson junctions (JJs) were employed instead of the standard Nb-based tri-layer JJs to improve the targeting precision of the designed critical current ($I_{\mathrm{c}}$).
    \item \textbf{Shunt Resistor:} Shunt resistors were composed of Ti/Pd (thickness, \SI[parse-numbers=false]{3.6/70}{\nano\metre}) to maintain stable shunt resistance ($R_{\mathrm{s}}$) control within the millikelvin temperature regime.
    \item \textbf{Dielectric Loss Mitigation:} Prior to the MCM process, the SiO$_2$ layer on the digital wafer was selectively removed via Buffered Oxide Etch (BOE) in areas proximal to the qubits to minimize dielectric-induced decoherence.
\end{itemize}

\subsection{Qubit chip}
Fabrication of the qubit chip utilized a \SI{2}{inch} c-cut sapphire substrate, with base-layer patterning mirroring the methods in Ref.~\cite{Fei2025}. Al/AlO$_x$/Al Josephson junctions and arrays were realized via Dolan bridge evaporation. Specifically, a \SI{1.3}{\micro\metre} LOR-B / \SI{300}{\nano\metre} PMMA-A4 bilayer resist was patterned using a Raith EBPG5200 system. Development consisted of \SI{2}{\minute} in MIBK/IPA (1:3), a \SI{1.5}{\minute} IPA rinse, and \SI{13}{\second} in MIF319. The junction deposition involved two Al layers (\SI{50}{\nano\metre} and \SI{150}{\nano\metre}) separated by oxidation at \SI{4}{Torr} for \SI{20}{\minute}. The process concluded with a \SI{3}{\hour} lift-off in \SI{70}{\celsius} NMP and standard solvent cleaning (acetone and IPA).

\subsection{MCM Integration}
\begin{figure}
    \includegraphics[width = \linewidth]{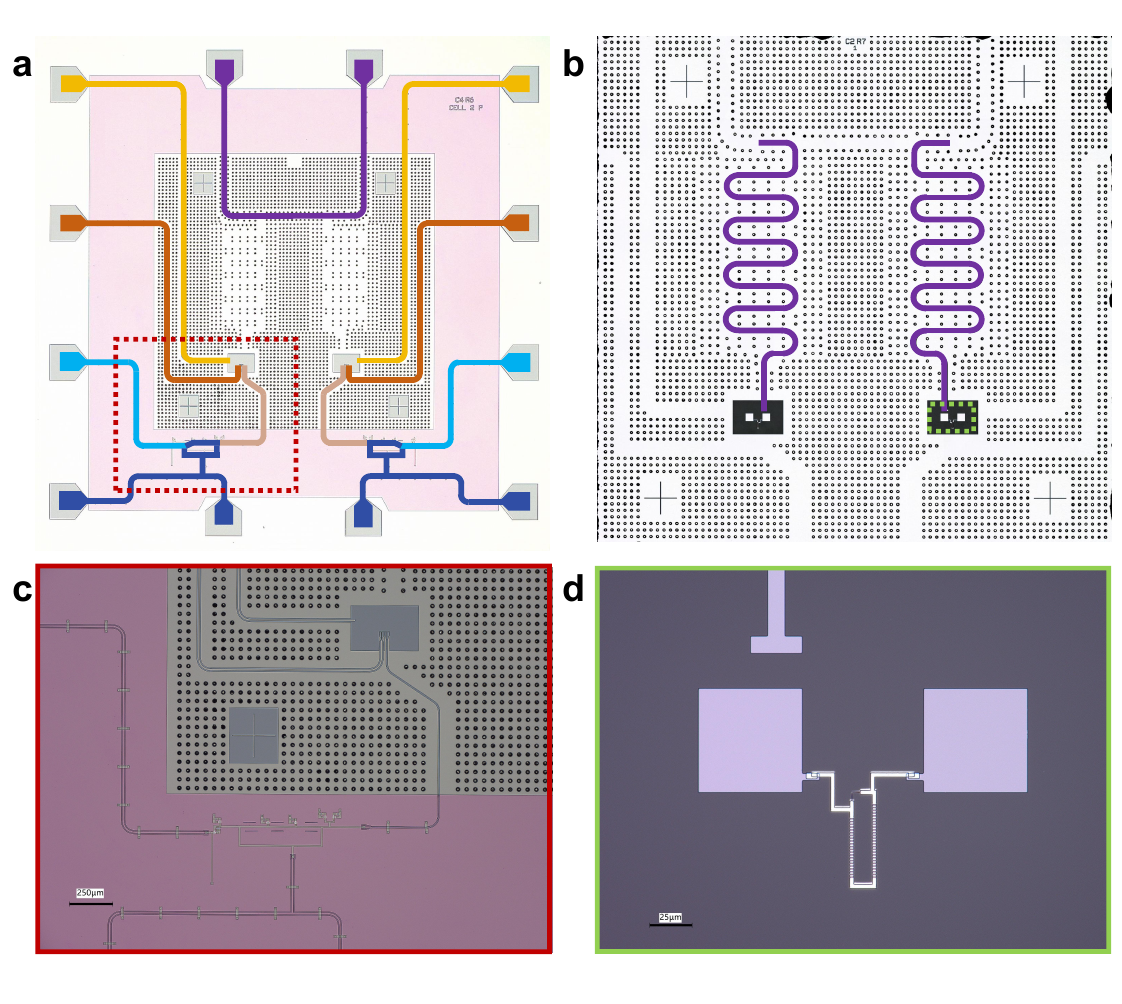} 
    \caption{\textbf{Optical micrographs of the fabricated chips.} \textbf{a,} False colored micrograph of the digital control chip; the left red box highlights the digitally controlled DAC device and the qubit control lines. \textbf{b,} False colored micrograph of the qubit chip. \textbf{c, d,} Magnified views of the control and the qubit structure, respectively. Color code: red for qubit flux lines, orange for qubit charge lines, and purple for the readout feedline (bottom chip) and resonators (top chip). Color coding for others (DAC) remains consistent with the schematic as shown in Fig.~1a of the main text.}
    \label{fig:microscope}
\end{figure}

For integration, \SI{12}{\micro\metre} thick Indium (In) bumps were deposited on both the qubit and digital wafers via thermal evaporation, followed by dicing into individual dies. The optical micrograph of the as-fabricated digital control chip is presented in \figref{fig:microscope}a and c, showing the full chip layout and the magnified device details, respectively. The as-fabricated qubit chip is displayed in \figref{fig:microscope}b and d, where the latter provides a magnified view of the fluxonium qubit structure. The two components were then bonded in a flip-chip configuration using a high-precision sub-micron aligner, maintaining a precisely controlled inter-chip gap of approximately \SI{10}{\micro\metre}. This vertical integration architecture ensures efficient inductive coupling between the DAC and the fluxonium qubit while providing a compact footprint suitable for future large-scale expansion.

\section{DAC parameters}\label{app:DACpara}

\begin{table}[ht]
\centering
\caption{Designed and measured (or derived) DAC parameters for different chips and qubits. The DAC storage inductance is designed to be $L = 1\,\mathrm{nH}$ for all devices. The derived critical current $I_c$ is estimated based on the process control monitoring of the junction critical current density $J_c$.} \label{tab:dac}
\begin{tabular}{lccccccccc}
\toprule
\multirow{2}{*}{Chip} & \multirow{2}{*}{DAC}
& \multicolumn{2}{c}{Range ($\Phi_0$)}
& \multicolumn{2}{c}{Step (m$\Phi_0$)}
& \multicolumn{2}{c}{$M$ (pH)}
& \multicolumn{2}{c}{$I_c$ ($\mu \mathrm{A}$)} \\
\cmidrule(lr){3-10}
& 
& Des. & Meas.
& Des. & Meas.
& Des. & Der.
& Des. & Der. \\
\midrule
C4R1 & 1 & 0.75 & 0.70 & 9.75 & 4.58 & 9.75 & 4.58 & 80 & 93 \\
     & 2 & 0.75 & 1.26 & 9.75 & 6.00 & 9.75 & 6.00 & 80 & 93 \\
\midrule
C1R5 & 1 & 0.35 & 0.25 & 4.60 & 3.70 & 4.60 & 3.70 & 80 & 93 \\
     & 2 & 0.35 & 0.35 & 4.60 & 4.70 & 4.60 & 4.70 & 80 & 93 \\
\midrule
C3R1 & 1 & 0.26 & 0.11 & 2.25 & $<1$ & 2.25 & $<1$ & 115 & 133 \\
     & 2 & 0.26 & 0.67 & 2.25 & 2.65 & 2.25 & 2.65 & 115 & 133 \\
\bottomrule
\end{tabular}
\end{table}

We characterize multiple flux DACs coupled to different qubits across several chips by measuring the shift in the qubit external flux as a function of the applied DAC bias. The maximum and minimum flux shifts directly determine the DAC tuning range, while the discrete flux shifts between successive DAC states define the effective step size. The extracted DAC range and step size for each device are summarized in \tabref{tab:dac}.

The DAC step size and total tuning range probe different aspects of the device. The (derived) step size equals to $M\Phi_0/L$, and its qubit-to-qubit variation therefore primarily reflects variations in the effective mutual inductance $M$, likely arising from chip alignment and coupling uncertainty in the MCM. The total DAC range equals to the step size multiplied by the number of accessible DAC states. Although the ideal accessible state number is set by $L I_c/\Phi_0$, the measured range can be reduced or distorted by flux trapping, asymmetric programming thresholds, and limitations of the monotonic operating window. Accordingly, the measured range is also governed by the fraction of usable DAC states under the chosen bias conditions.

\section{RCSJ model}\label{app:RCSJ}

The temporal evolution of the DAC with external flux disturbances can be captured by the RCSJ model ~\cite{SQUIDHandbook2004}:
\begin{equation*}
   \beta_c \frac{d^2 \phi}{d\tau^2} + \frac{d\phi}{d\tau} + \sin\phi = \frac{1}{\beta_L}(\phi - \phi_{\text{ext}})
\end{equation*}
where $\phi= 2\pi\Phi/\Phi_0$, $\phi_\mathrm{ext} =2\pi\Phi_\mathrm{ext}/\Phi_0$, $\tau = \omega_c t$ with the Josephson characteristic frequency $\omega_c = 2\pi I_c R/\Phi_0$. The final state of the DAC after removing the external flux disturbance is determined by the Stewart–McCumber parameter $\beta_c = 2\pi I_c R^2 C/\Phi_0$. Benefiting from the shunt resistor process described in \appref{app:fab}, our device has a specific capacitance of $C = 3.5 \,\mathrm{pF}$, a normal-state resistance of $R_n = 3.5 \,\Omega$, a shunt resistor of $R_\mathrm{shunt}=0.45\,\Omega$ and a critical current of $I_c = 80 \,\mu\mathrm{A}$. Consequently, $\beta_c$ is reduced from 10.4 to 0.135, which indicates that Josephson junction changes from an under-damped to a strongly overdamped regime. In the strongly overdamped limit, DAC will relax to and remain in the first metastable state encountered during process $\textcircled{\small 2}$ as shown in Fig.~2b of the main text, so that it allows for stepwise tuning of the DAC within its operating range $\pm LI_0\sim\pm38\,\Phi_0$.

\section{Experimental setup}\label{app:exp}

\begin{figure}
    \includegraphics[width = \linewidth]{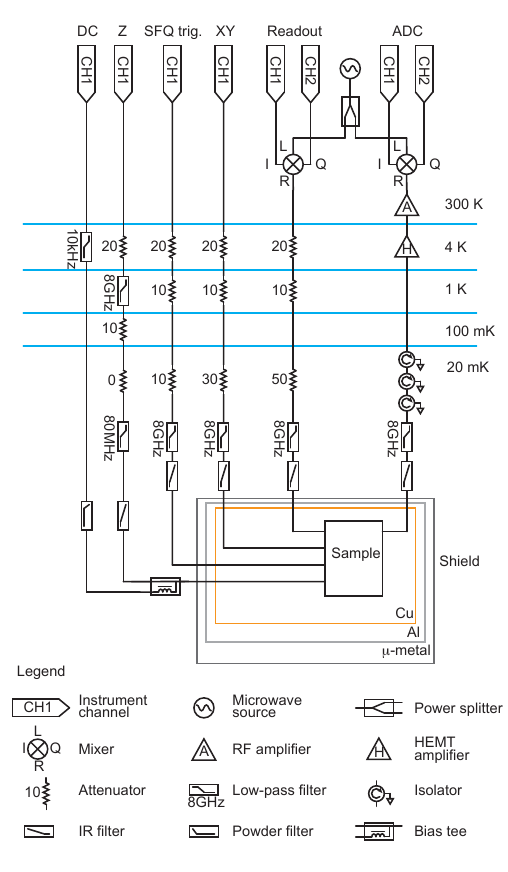} 
    \caption{\textbf{Schematics of the experimental setup.}}
    \label{fig:meas_setup}
\end{figure}

The qubit is calibrated in a dilution refrigerator at a base temperature below 20 mK. The setup is similar with those in Ref.~\cite{Fei2025}. The total attenuation in flux line/SFQ bias (Z, as depicted in \figref{fig:meas_setup}) has been reduced to 30 dB and the additional 25 MHz low pass filter (LPF) has been removed. In addition, we removed 20 dB attenuation on the 10 mK plate of an XY line, which serves as the SFQ trigger.

For the room-temperature electronics, we use home-made arbitrary waveform generators (AWGs) with a sampling rate of 2 GS/s to output control waveforms for the qubits, DAC, and dc/SFQ convertor. The readout circuit is modulated and demodulated using a pair of IQ mixers with respective local oscillators (LOs), and the readout signals are also obtained via home-made analog-to-digital converters (ADCs).

\section{DAC Programming protocols} \label{app:protocol}

\begin{figure}
    \includegraphics[width = \linewidth]{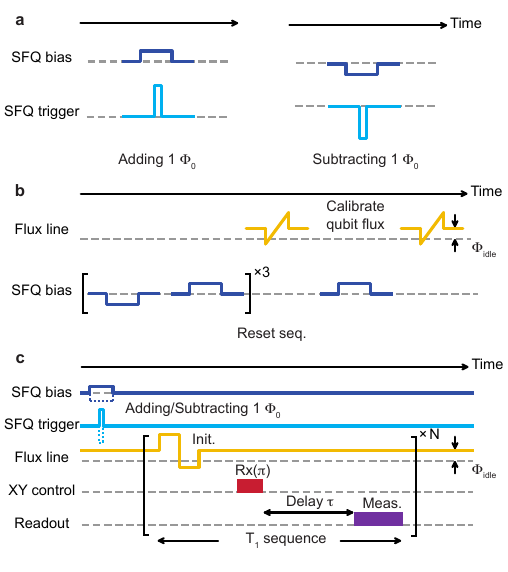} 
    \caption{\textbf{DAC Programming protocols.} \textbf{a,} Protocol for adding or subtracting one flux quantum $\Phi_0$. The gray dashed lines indicate the zero level of the corresponding channels.  
    \textbf{b,} Reset protocol for initializing the DAC to a deterministic reference state. \textbf{c,} A typical qubit experiment following DAC programming. The qubit flux is increased (solid line) or decreased (dashed line) by a single SFQ bias and trigger square pulse. The energy relaxation time $T_1$ at the corresponding flux bias is then measured using conventional techniques with $N$ repetitions.}
    \label{fig:dac_protocol}
\end{figure}

The DAC is programmed using an SFQ-based control chain consisting of a bias line and a trigger line. As shown in Fig.~\ref{fig:dac_protocol}a, controlled updates of the DAC state are achieved by applying synchronized square pulses to these lines, generating SFQ pulses that induce transitions between adjacent metastable states. Each pulse deterministically adds or subtracts one flux quantum $\Phi_0$ in the DAC storage loop, enabling stepwise and programmable control of the DAC output.

To ensure reproducible operation, the DAC is initialized to a known reference state prior to programming. As illustrated in Fig.~\ref{fig:dac_protocol}b, a reset pulse sequence is applied to bring the DAC to a deterministic initial state. This is achieved by alternately saturating the DAC to its minimum and maximum states multiple times, ensuring reliable convergence to a well-defined starting point.

The analog output of the DAC is characterized through its effect on a flux-tunable qubit. As shown in Fig.~\ref{fig:dac_protocol}c, an external analog flux bias $\Phi_\mathrm{idle}$ is applied to the qubit, allowing the DAC-induced flux shift to be measured. By comparing the qubit response before and after each DAC programming step, the effective flux increment per step can be extracted, enabling calibration of the DAC output.

\section{Qubit relaxation time}\label{app:T1}
\begin{figure}
    \includegraphics[width=1\linewidth]{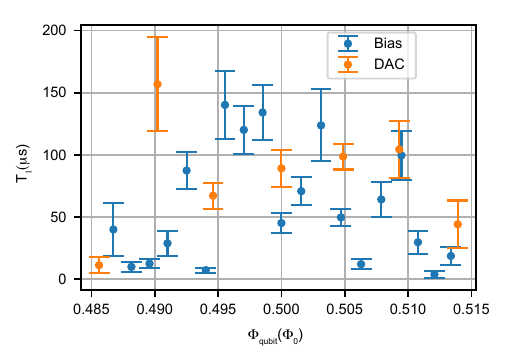}
    \caption{\textbf{$T_1$ versus external flux for fluxonium under conventional bias and DAC control.} Blue markers show measurements obtained using the conventional flux bias line, while orange markers correspond to flux tuning via the flux DAC.}
    \label{fig:T1}
\end{figure}

We measure the qubit relaxation time $T_1$ as a function of the qubit external flux $\Phi_{\mathrm{qubit}}$ by sweeping the conventional flux bias and the DAC output.

As shown in Fig.~\ref{fig:T1}, $T_1$ is characterized over the same flux range as $T_2$ around the sweet spot ($\Phi_{\mathrm{qubit}} \approx 0.5\,\Phi_0$). The average measured $T_1$ in this range is approximately $58\,\mu\mathrm{s}$ for conventional bias and $81\,\mu\mathrm{s}$ under DAC control.

Overall, both control schemes exhibit similar distributions of $T_1$. The general agreement indicates that DAC-based flux control does not introduce additional dissipation or relaxation channels.

\section{Scalable DAC addressing scheme}\label{app:scale}
\begin{figure}
    \includegraphics[width=1\linewidth]{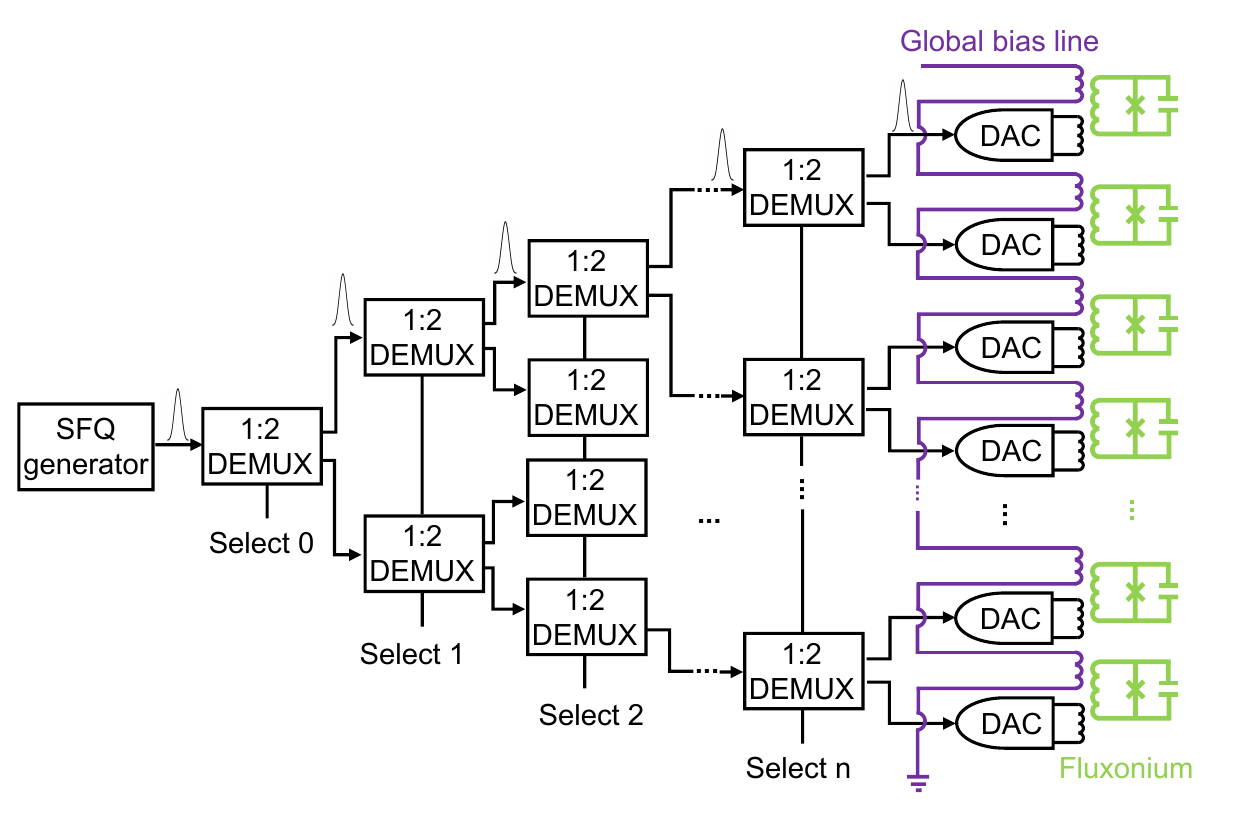}
    \caption{\textbf{Schematic of a scalable DAC programming architecture.} The control logic employs an SFQ generator followed by a multi-level 1:2 DEMUX tree, with each stage controlled by addressing lines (Select $0, 1, \dots, n$) to route digital signals to the target DAC. A shared bias line provides a global coarse flux offset for all qubits and serves as a reference for DAC calibration through the qubits.}
    \label{fig:demux}
\end{figure}

To enable scalable programming of a large number of flux DACs, we propose an SFQ-based demultiplexing (DEMUX) addressing architecture, as illustrated in Fig.~\ref{fig:demux}. SFQ pulses generated from a common source are routed through a binary tree of 1:2 DEMUX stages, where each stage is controlled by a low-bandwidth select line. In this scheme, $n$ DACs can be individually addressed using only $\log_2(n)$ control lines. 

In parallel, qubit biasing is implemented using a combination of a shared coarse analog bias and local fine adjustment from the flux DACs. A common bias line provides a global coarse flux offset for all qubits, while each flux DAC generates local flux to fine-tune the operating point of its corresponding qubit. Because the required flux bias across qubits typically lies within a narrow range, the global coarse bias reduces the output range needed for each DAC. In addition, the same global bias serves as a reference for DAC calibration through qubit-based measurements.

\bibliography{refs}